\documentclass[twocolumn,showpacs,preprintnumbers,amsmath,amssymb]{revtex4-1}
\usepackage{graphicx}
\usepackage{rotating}
\begin{document}

\renewcommand{\theequation}{\thesection.\arabic{equation}}

\newcommand{\re}{\mathop{\mathrm{Re}}}

\newcommand{\be}{\begin{equation}}
\newcommand{\ee}{\end{equation}}
\newcommand{\bea}{\begin{eqnarray}}
\newcommand{\eea}{\end{eqnarray}}


\title{Redshift drift in varying speed of light cosmology}

\author{Adam Balcerzak}
\email{abalcerz@wmf.univ.szczecin.pl}
\affiliation{\it Institute of Physics, University of Szczecin, Wielkopolska 15, 70-451 Szczecin, Poland}
\affiliation{\it Copernicus Center for Interdisciplinary Studies, S{\l }awkowska 17, 31-016 Krak\'ow, Poland}

\author{Mariusz P. D\c{a}browski}
\email{mpdabfz@wmf.univ.szczecin.pl}
\affiliation{\it Institute of Physics, University of Szczecin, Wielkopolska 15, 70-451 Szczecin, Poland}
\affiliation{\it Copernicus Center for Interdisciplinary Studies,
S{\l }awkowska 17, 31-016 Krak\'ow, Poland}

\date{\today}

\input epsf

\begin{abstract}
We derive a redshift drift formula within the framework of varying speed of light (VSL) theory using the specific ansatz for the variability of $c(t) = c_0 a^n(t)$. We show that negative values of the parameter $n$, which correspond to diminishing value of the speed of light during the evolution of the universe, effectively
rescales dust matter to become little negative pressure matter, and the cosmological constant to became phantom. Positive values of $n$ (growing $c(t)$) make VSL model to become more like Cold Dark Matter (CDM) model. Observationally, there is a distinction between the VSL model and the $\Lambda$CDM model for the admissible values of the parameter $n \sim - 10^{-5}$, though it will be rather difficult to detect by planned extremely large telescopes (E-ELT, TMT, GMT) within their accuracy.
\end{abstract}


\maketitle

\section{Introduction}
\label{intro}
\setcounter{equation}{0}

The early idea of variation of physical constants \cite{varconst} has been established widely in physics both theoretically and experimentally \cite{uzan}. The gravitational constant $G$, the charge of electron $e$, the velocity of light $c$, the proton to electron mass ratio $\mu = m_p/m_e$, and the fine structure constant $\alpha = e^2/\hbar c$, where $\hbar$ is the Planck constant, may vary in time and space \cite{barrowbook}. The earliest and best-known framework for varying $G$ theories has been Brans-Dicke theory \cite{bd}. Nowadays, the most popular theories which admit physical constants variation are the varying $\alpha$ theories \cite{alpha}, and the varying speed of light $c$ theories \cite{uzanLR,VSL}. The latter, which will be the interest of our paper, allow the solution of the standard cosmological problems such as the horizon problem, the flatness problem, the $\Lambda-$problem, and has recently been proposed to solve the singularity problem \cite{JCAP13}.

Recently, lots of interest has been attracted by the effect of redshift drift in cosmological models. This effect was first noticed by Sandage and later explored by Loeb \cite{sandage+loeb}. The idea is to collect data from the two light cones separated by the time period of 10-20 years to look for the change of redshift of a source in time $\Delta z/\Delta t$ as a function of redshift of this source. The effect has recently been investigated for the inhomogeneous density Lemaitre-Tolman-Bondi models \cite{LTB,UCETolman}, the Dvali-Gabadadze-Porrati (DGP) brane model \cite{Quercellini12}, backreaction timescape cosmology \cite{wiltshire}, axially symmetric Szekeres models \cite{marieN12}, inhomogeneous pressure Stephani models \cite{PRD13}. In Ref.\cite{Quercellini12} the drift for the $\Lambda$CDM model, the Dvali-Gabadadze-Porrati (DGP) brane model, the matter-dominated model (CDM), and three different LTB void models have been presented. It has been shown that the drift for $\Lambda$CDM and DGP models is positive up to $z \approx 2$ and becomes negative for larger redshifts, while it is always negative for LTB void models \cite{LTB,yoo}. The drift for Stephani models becomes positive for small redshifts and approaches the behavior of the $\Lambda$CDM model, which allows negative values of the drift, for very high redshifts \cite{PRD13}. The effect of varying constants theories including VSL theories onto the redshift drift has not yet been investigated and that is the motivation for this work.

Our paper is organized as follows. In Sec. \ref{VSL} we formulate the basics of the varying speed of light (VSL) theory and define
observational parameters such as the dimensionless energy density parameters $\Omega$, Hubble parameter $H$, deceleration parameter $q$, as well as the higher
derivative parameters like jerk $j$, snap $s$ etc. \cite{jerk,snap,weinberg} which may serve as indicators of the equation of state (statefinders)
and the curvature of the universe. In Sec. \ref{drift} we derive the redshift drift formula for the VSL cosmology using special ansatz for the time dependence of the speed of light $c(t) = c_0 a^n(t)$, where $a(t)$ is the scale factor and $c_0$, $n$ are constants. In Sec. \ref{conclusion} we give our conclusions.

\section{Varying speed of light theory}
\setcounter{equation}{0}
\label{VSL}
Following the Ref. \cite{VSL}, we consider the Friedmann universes within the framework of varying speed of light theories (VSL) with the metric
\be
ds^2 = - (dx^0)^2 + a^2(t) \left[ \frac{dr^2}{1-Kr^2} + r^2 (d\theta^2 + \sin^2{\theta} d\varphi^2) \right]~~,
\ee
where $dx^0 = c(t) dt$, for which the field equations read
\bea \label{rho} \varrho(t) &=& \frac{3}{8\pi G}
\left(\frac{\dot{a}^2}{a^2} + \frac{Kc^2(t)}{a^2}
\right)~,\\
\label{p} p(t) &=& - \frac{c^2(t)}{8\pi G} \left(2 \frac{\ddot{a}}{a} + \frac{\dot{a}^2}{a^2} + \frac{Kc^2(t)}{a^2} \right)~,
\eea
and the energy-momentum conservation law is
\be
\label{conser}
\dot{\varrho}(t) + 3 \frac{\dot{a}}{a} \left(\varrho(t) + \frac{p(t)}{c^2(t)} \right) = 3 \frac{Kc(t)\dot{c}(t)}{4\pi Ga^2}~.
\ee
Here $a \equiv a(t)$ is the scale factor, the dot means the derivative with respect to time $t$, $G$ is the gravitational constant, $c=c(t)$ is time-varying speed of light, and the curvature index $K=0, \pm 1$. In most of the paper we will follow the ansatz for the speed of light given in Ref. \cite{BM99}, i.e.,
\be
\label{c(t)}
c(t) = c_0 a^n(t)~~,
\ee
with the constant speed of light limit $n \to 0$ giving $c(t) \to c_0$. We have $\dot{c}/c = n \dot{a}/a$, so the speed of light grows in time for $n>0$, and diminishes for $n<0$.

The cosmological observables which characterize the kinematic evolution of the universe are \cite{PLB05}:
\\
\noindent the Hubble parameter
\be
\label{hubb}
H = \frac{\dot{a}}{a}~,
\ee
the deceleration parameter
\be
\label{dec}
q  =  - \frac{1}{H^2} \frac{\ddot{a}}{a} = - \frac{\ddot{a}a}{\dot{a}^2}~,
\ee
the jerk parameter \cite{jerk}
\be
\label{jerk}
j = \frac{1}{H^3} \frac{\dddot{a}}{a} =
\frac{\dddot{a}a^2}{\dot{a}^3}~,
\ee
and the snap \cite{snap} parameter
\be
\label{snap}
s = -\frac{1}{H^4} \frac{\ddddot{a}}{a} =
-\frac{\ddddot{a} a^3}{\dot{a}^4}~.
\ee
We can carry on with these and define even the higher derivative parameters
such as lerk (crack), merk (pop), etc. \cite{PLB05,ANN06,gibbons} by
\be
x^{(i)} = (-1)^{i+1} \frac{1}{H^{i}} \frac{a^{(i)}}{a} = (-1)^{i+1} \frac{a^{(i)} a^{i-1}}{\dot{a}^{i}}~~,
\ee
where $i = 2, 3, ...$, and $a^{(i)}$ means the i-th derivative with respect to time while $a^{i}$ means the n-th power. We have consecutively:
$q$ for $i=2$, $j$ for $i=3$ etc.

A comparison of cosmological models with observational data requires the introduction of dimensionless density parameters
\bea
\label{Omegadef}
\Omega_{m0}  &=&  \frac{8\pi G}{3H_0^2} \varrho_{m0} ,\\
\Omega_{K0}  &=&  \frac{Kc_0^2a_0^{2n}}{H_0^2a_0^2} ,\\
\Omega_{\Lambda_0}  &=&  \frac{\Lambda_0 c_0^2a_0^{2n}}{3H_0^2} ,
\eea
for dust, curvature, and dark energy, respectively. The index "0" means that we take these parameters at the present
moment of the evolution $t=t_0$.

The following relations are valid \cite{ANN06}
\bea
\label{Omegarel}
\Omega_{K0} &=& \frac{3}{2} \Omega_{m0} - (q_0 + n) - 1 ,\\
j_0 &=& \Omega_{m0} + \Omega_{\Lambda 0}\left( n+1 \right) - n \Omega_{K0},\\
\Omega_{\Lambda 0}\left( n+1 \right) &=& \frac{1}{2} \Omega_{m0} - q_0 + n \Omega_{K0},
\eea
and so
\be
\label{j0rel}
j_0 + 1 + \Omega_{K0} = 3 \Omega_{m0} - 2 q_0 - n.
\ee

\section{Redshift drift in varying speed of light theory}
\setcounter{equation}{0}
\label{drift}

\begin{figure}[ht]
 \begin{center}
 \scalebox{1.0}{\includegraphics[angle=0]{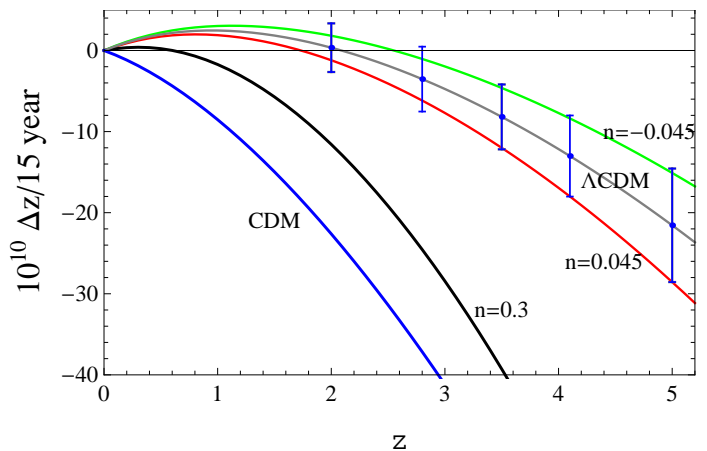}}
\caption{The redshift drift effect (\ref{delzomega}) for 15 year period of observations for various values of the varying speed of light parameter $n$. Negative $n$ correponds to $\dot{c} < 0$. The error bars are taken from Ref. \cite{Quercellini12} and presumably show that for $|n| < 0.045$ one cannot distinguish between VSL models and $\Lambda$CDM models. Larger positive values of $n$ (growing $c(t)$) make VSL model to become more like Cold Dark Matter (CDM) dust model.}
\label{fig1}
 \end{center}
 \end{figure}

We consider redshift drift effect in VSL theory. In order to do that we assume that the source does not possess any peculiar velocity, so that it maintains a fixed comoving coordinate $dr=0$. The light emitted by the source at two different moments of time $t_e$ and $t_e+\delta t_e$ in VSL universe will be observed at $t_o$ and $t_o+\delta t_o$ related by
\be
\int_{t_e}^{t_o}\frac{c(t) dt}{a(t)}=\int_{t_e+\Delta t_e}^{t_o+\Delta t_o}\frac{c(t) dt}{a(t)}~,
\ee
which for small $\Delta t_e$ and $\Delta t_o$ transforms into
\be
\label{rel}
\frac{c(t_e) \Delta t_e}{a(t_e)}=\frac{c(t_0) \Delta t_o}{a(t_o)}~.
\ee
The definition of redshift in VSL theories remains the same as in standard Einstein relativity and reads as \cite{BM99}
\be
\label{redshift}
1+z = \frac{a(t_0)}{a(t_e)}~~.
\ee
The redshift drift is defined as \cite{sandage+loeb}
\begin{eqnarray}
\label{redshiftdrift}
\Delta z = z_e - z_0 = \frac{a(t_0 + \Delta t_0)}{a(t_e + \Delta t_e)} - \frac{a(t_0)}{a(t_e)}~,
\end{eqnarray}
which can be expanded in series (cf. Appendix) and to first order in $\Delta t$ reads as
\bea
&& \Delta z = \frac{a(t_0) + \dot{a}(t_0)\Delta t_0}{a(t_e) + \dot{a}(t_e)\Delta t_e} - \frac{a(t_0)}{a(t_e)} \nonumber \\
&& \approx \frac{a(t_0)}{a(t_e)} \left[ \frac{\dot{a}(t_0)}{a(t_0)} \Delta t_0 - \frac{\dot{a}(t_e)}{a(t_e)} \Delta t_e \right]~~.
\eea
Using (\ref{rel}) we have
\be
\label{delz1}
\Delta z = \Delta t_0 \left[ H_0 (1+z) - H(t_e) \frac{c(t_0)}{c(t_e)} \right]~~,
\ee
which after applying the ansatz (\ref{c(t)}) gives
\be
\label{delz2}
\frac{\Delta z}{\Delta t_0} = \frac{\Delta z}{\Delta t_0} (z,n) = H_0 (1+z) - H(z) (1+z)^{n}~~.
\ee
In the limit $n \to 0$ the formula (\ref{delz2}) reduces to the standard constant speed of light Friedmann universe
formula obtained by Sandage and Loeb \cite{sandage+loeb}. Bearing in mind the definitions $\Omega$'s and assuming $K=0$ we have
\be
H^2(z) = H_0^2 \left[ \Omega_{m0} (1+z)^3 + \Omega_{\Lambda} \right]
\ee
and so (\ref{delz2}) gives
\bea
\label{delzomega}
&& \frac{\Delta z}{\Delta t_0} = H_0 \left[ 1+z - (1+z)^n \sqrt{\Omega_{m0} (1+z)^3 + \Omega_{\Lambda}} \right] \nonumber \\
 &=& H_0 \left[ 1+z - \sqrt{\Omega_{m0} (1+z)^{3+2n} + \Omega_{\Lambda}(1+z)^{2n}} \right]
\eea
which can further be rewritten to define new redshift function
\be
\tilde{H}(z) \equiv (1+z)^n H(z) = H_0 \sqrt{\sum_{i=1}^{i=k} \Omega_{wi} (1+z)^{3(w_{eff} + 1)}}~~,
\ee
where
\be
w_{eff} = w_i + \frac{2}{3} n~~.
\ee
Using (\ref{delzomega}) we present a plot of the redshift drift in VSL models in Fig. \ref{fig1}. For the negative values of the parameter $n$ which correspond to diminishing value of the speed of light during the evolution of the universe, it effectively rescales dust matter to become little negative pressure matter, and the cosmological constant to became phantom \cite{phantom}. In other words, both components become extra sources of dark energy. Positive values of $n$ (growing $c(t)$) make VSL model to become more like Cold Dark Matter (CDM) model. Then, both matter components (dust, cosmological term) become extra sources of dark energy for $n \sim - 10^{-5} <0$ which is in agreement with observational data \cite{murphy2007,king2012}. In Fig. \ref{fig1} the theoretical error bars are taken from Ref. \cite{Quercellini12} and presumably show that for $|n| < 0.045$ one cannot distinguish between VSL models and $\Lambda$CDM models. However, if the bars are reduced, then the influence of varying $c$ onto the evolution of the universe may perhaps be distinguishable.

Different predictions for redshift drift in various cosmological models can be tested in future telescopes such as the European Extremely Large Telescope (EELT) (with its spectrograph CODEX (COsmic Dynamics EXperiment)) \cite{balbi,E-ELT}, the Thirty Meter Telescope (TMT), the Giant Magellan Telescope (GMT), and especially, in gravitational wave interferometers DECIGO/BBO (DECi-hertz Interferometer Gravitational Wave Observatory/Big Bang Observer) \cite{DECIGO}. The first class of the experiments involving the very sensitive spectrographic techniques such as those utilized in the CODEX spectrograph use a detection of a very slow time variation of the Lyman-$\alpha$ forest of the number of quasars uniformly distributed all over the sky to measure the redshift drift, but  Lyman-$\alpha$ lines become impossible to measure for  $z<1.7$ from the ground \cite{E-ELT}. The lower range of redshifts can be investigated though in other class of future experiments involving the space-borne gravitational wave interferometers DECIGO/BBO \cite{DECIGO}. A detection could be possible even at $z\sim 0.2$.

\section{Conclusions}
\setcounter{equation}{0}
\label{conclusion}

We have calculated a redshift drift formula in varying speed of light theory. The formula is valid for any time dependence of the velocity of light though we have used  the specific ansatz for the variability of $c(t) = c_0 a^n(t)$ in order to discuss the effect of varying $c$ onto the redshift change over the evolution of the universe.
We have shown that for observationally admissible negative values of the parameter $n \sim - 10^{-5} < 0$ ($\dot{c}(t) <0$) all the components of the universe behave as extra sources of dark energy. On the other hand, positive values of $n$ ($\dot{c}(t)>0$) make VSL models to decelerate and behave more like Cold Dark Matter (CDM) models.

By using the theoretical error bars from Ref. \cite{Quercellini12} we have shown (cf. Fig. \ref{fig1}) that for $|n| < 0.045$ one basically cannot distinguish between VSL models and $\Lambda$CDM models. However, if the bars are reduced, then the influence of varying $c$ onto the evolution of the universe may perhaps be distinguishable.
In any case, the redshift drift will become an independent test of the VSL universe since it potentially shows the difference from the $\Lambda$CDM universe.

The potential detection of the effect of redshift drift will be possible by extremely large telescopes such as EELT, TMT, and GMT. There is also some hope that these experiments give better accuracy in space-born future gravitational wave detectors such as DECIGO/BBO.

It is worth mentioning that our derivation of redshift drift formula (\ref{delz2}) would even fit better the prospective data, if the ansatz $c(t) = c_0 a^{n(t)}$ of Ref. \cite{BM99} was applied. With such a variable $n$ parameter ansatz, one would be able to match the variability of $c$ with the cosmic evolution following the suggestion of \cite{BM99} in the sense that $n$ was larger ($n = -2.2$) in the radiation epoch, and then it was gradually diminishing to reach the value
$n \sim - 10^{-5} < 0$ which is compatible with the current observational constraints on $c \propto {\alpha}^{-1}$ \cite{murphy2007,king2012}.

\section{Acknowledgements}

This project was financed by the National Science Center Grant DEC-2012/06/A/ST2/00395.

\appendix

\section{Higher-order statefinder redshift drift formula}

The scale factor $a(t)$ at any moment of time $t$ can be obtained as series expansion around $t_0$ as ($a(t_0) \equiv a_0$) \cite{ANN06}
\bea
\label{seriesa}
&&a(t) = a_0 \left\{ 1 + H_0 (t-t_0) - \frac{1}{2!}q_0 H_0^2
(t-t_0)^2 \right.  \\
&& \left. + \frac{1}{3!} j_0 H_0^3 (t-t_0)^3 - \frac{1}{4!} s_0
H_0^4 (t-t_0)^4 + O[(t-t_0)^5]\right\}~, \nonumber
\eea
and its inverse reads as
\bea
\label{seriesinva}
&&\frac{a_0}{a(t)} = 1+z = 1 + H_0 (t_0-t) + H_0^2
\left(\frac{q_0}{2} +1 \right) (t_0-t)^2 \nonumber \\
&& + H_0^3 \left(q_0 + \frac{j_0}{6} + 1 \right) (t_0-t)^3 \\
&& + H_0^4 \left(1 + \frac{j_0}{3} + \frac{q_0^2}{4} + \frac{3}{2} q_0
+ \frac{s_0}{24} \right) (t_0-t)^4 \nonumber \\
&& + O[(t_0-t)^5]~. \nonumber
\eea

Using (\ref{seriesa}) and (\ref{seriesinva}), the redshift drift formula (\ref{redshiftdrift}) can be expanded up to higher order characteristics of the expansion $q_0$, $j_0$, and $s_0$ as
\bea
&& \Delta z = \frac{a(t_0)}{a(t_e)} \left[ {\bf H_0 \Delta t_0 - H_e \Delta t_e} - H_0 H_e \Delta t_0 \Delta t_e \right. \nonumber \\
&& \left. - \frac{1}{2} q_0 H_0^2 (\Delta t_0)^2 + H_e^2 \left( \frac{q_e}{2} + 1 \right) (\Delta t_e)^2 \right. \nonumber \\
&& \left. + \frac{1}{3!} j_0 H_0^3 (\Delta t_0)^3 - H_e^3 \left( \frac{j_e}{3} + q_e + 1 \right) (\Delta t_e)^2 \right. \nonumber \\
&& \left. + H_0 H_e^2 \left( \frac{q_e}{2} + 1 \right) (\Delta t_0)(\Delta t_e)^2 \right. \nonumber \\
&& \left. + \frac{1}{2} q_0 H_0^2 H_e^2 \left( \frac{q_e}{2} + 1 \right) (\Delta t_0)^2 (\Delta t_s)^2 \right. \nonumber \\
&& \left. - \frac{1}{4} s_0 H_0^4 (\Delta t_0)^4 + H_e^4 \left( 1 + \frac{j_e}{3} + \frac{q_e^2}{4} + \frac{3}{2} q_e + \frac{s_e}{24} \right) \right. \nonumber \\
&& \left. + (\Delta t_e)^4 + O \left[(\Delta t)^5 \right] \right]~~,
\eea
where only the first two terms appear in the first order formula (\ref{redshiftdrift}).

\end{document}